\documentclass[12pt]{article} 
\usepackage{amssymb}
\usepackage{graphicx}
\begin{document}  
\begin{center}
{\large \bf Critical behavior of cross sections at LHC}

\vspace{0.5cm}                   

{\bf I.M. Dremin$^{1,2}$}

\vspace{0.5cm}                       

        $^1$Lebedev Physics Institute, Moscow 119991, Russia\\
\medskip

    $^2$National Research Nuclear University "MEPhI", Moscow 115409, Russia     

\end{center}

\begin{abstract}
Recent experimental data on elastic scattering of high energy protons show
that the critical regime has been reached at LHC energies. The approach to 
criticality is demonstrated by increase of the ratio of elastic to total 
cross sections from ISR to LHC energies. At LHC it reaches the value which 
can result in principal change of the character of proton interactions. The 
treatment of new physics of hollowed toroid-like hadrons requires usage of 
another branch of the unitarity condition. Its further fate is speculated and 
interpreted with the help of the unitarity condition in combination with 
present experimental data. The gedanken experiments to distinguish between 
different possibilities are proposed.
\end{abstract}

\section{Introduction}

Recent experimental data on elastic scattering of high energy protons 
\cite{totem1, totem2} show quite surprising phenomenon of increase of the ratio 
of elastic to total cross sections with energy increase in the interval from 
ISR to LHC energies. This share used to decrease at lower energies but reversed 
the tendency at ISR (for the comparison see the tables in \cite{ufnel, drem2}). 
Moreover, at LHC energies it approaches the critical value \cite{jetp14, igse}. 
For the first time, that phenomenon can reveal the transition from the branch 
of the unitarity condition dominated by inelastic processes where elastic 
scattering is treated as the shadow of inelastic collisions to the dominance of 
elastic scattering which would require new interpretation. Elastic scattering 
of polarised protons or charge asymmetries of pions produced in inelastic 
collisions could help in studies of different possibilities. 

The information about elastic scattering comes from the measurement of the 
differential cross section $d\sigma /dt$ at some energy $s$ as a function of
the transferred momentum $t$ at its experimentally available values.
It is related to the scattering amplitude $f(s,t)$ in a following way
\begin{equation}
\frac {d\sigma }{dt}=\vert f(s,t)\vert ^2.
\label{dsdt}
\end{equation}
The variables $s$ and $t$ are the squared energy $E$ and transferred momentum 
of colliding protons in the center of mass system $s=4E^2=4(p^2+m^2)$, 
$-t=2p^2(1-\cos \theta)$ at the scattering angle $\theta $. From this 
measurement one gets the knowledge only about the modulus of the amplitude. 
The interference between the nuclear and Coulomb contributions to the amplitude 
$f$ allows to find out the ratio of the real and imaginary parts of the elastic 
scattering amplitude $\rho (s,t)={\rm Re}f(s,t)/{\rm Im}f(s,t)$ just in forward 
direction $t=0$ $\rho (s,0)=\rho _0$ but not at any other values of $t$.

The typical shape of the differential cross section at high energies contains
the exponentially decreasing (with increase of $\vert t\vert $) diffraction 
cone with energy dependent slope $B(s)$ and more slowly decreasing tail at 
larger transferred momenta with much smaller values of the cross section.

\section{The unitarity condition}

The most stringent and reliable information about the amplitude $f$ comes 
from the unitarity of the $S$-matrix 
\begin{equation}
SS^+=1
\label{ss}
\end{equation}
or for the scattering matrix $T \; (S=1+iT)$
\begin{equation}
2{\rm Im}T_{ab}=\Sigma _n\int T_{an}T^*_{nb}d\Phi _n,
\label{tt}
\end{equation}
where the whole $n$-particle phase space $\Phi _n$ is integrated over.
It relates the amplitude of elastic scattering $f=T_{22}$ to the amplitudes of 
$n$-particle inelastic processes $T_{2n}$ declaring that the total probability 
of all outcomes of the interaction must be equal 1\footnote{The non-linear
contribution from the elastic amplitude appears in the right-hand side for
$n=2$.}. In the $s$-channel this indubitable condition is usually
expressed in the form of the well known integral relation (for 
more details see, e.g., \cite{PDG, anddre1, ufnel}). This relation
can be simplified to the algebraic one using the Fourier -- Bessel transform 
of the amplitude $f$ which retranslates the momentum data to the shortest 
transverse distance between the trajectories of the centers of colliding 
protons called impact parameter $b$ and is written as
\begin{equation}
i\Gamma (s,b)=\frac {1}{2\sqrt {\pi }}\int _0^{\infty}d\vert t\vert f(s,t)
J_0(b\sqrt {\vert t\vert }).
\label{gamm}
\end{equation}
The unitarity condition in the $b$-representation reads (for the review see,
e.g., Refs \cite{ufnel, drem2})
\begin{equation}
G(s,b)=2{\rm Re}\Gamma (s,b)-\vert \Gamma (s,b)\vert ^2.
\label{unit1}
\end{equation}
The left-hand side describes the transverse impact-parameter profile of 
inelastic collisions of protons (for more detailed discussion see 
\cite{drem2, igse}). It satisfies 
the inequalities $0\leq G(s,b)\leq 1$ and determines how absorptive is the 
interaction region depending on the impact parameter (with $G=1$ for the full 
absorption and $G=0$ for the complete transparency). The profile of elastic 
processes is determined by the subtrahend in Eq. (\ref{unit1}). If $G(s,b)$ is
integrated over the impact parameter, it leads to the cross section of
inelastic processes. The terms on the right-hand side would produce the total
cross section and the elastic cross section, correspondingly.

\section{Central collisions}

At the beginning, let us study the energy dependence of interaction profiles
for central collisions of impinging protons at $b=0$. Then the condition
(\ref{unit1}) is written as 
\begin{equation}
G(s,b=0)= \zeta (2-\zeta),
\label{gZ}       
\end{equation}
 where
\begin{equation}
\zeta (s)=\frac {\sigma _{tot}(s)}{4\pi B(s)}= \frac {4\sigma _{el}(s)}
{(1+\rho _0 ^2(s))\sigma _{tot}(s)} \approx \frac {4\sigma _{el}(s)}
{\sigma _{tot}(s)}\approx  (4\pi )^{-0.5}
\int _0^{\infty}d\vert t\vert \sqrt {d\sigma /dt}.
\label{ze}
\end{equation}
is proportional to the experimentally measurable dimensionless ratio of the 
elastic cross section $\sigma _{el}$ (or 
the diffraction cone slope $B$) to the total cross section $\sigma _{tot}$.
It is computed integrating the experimentally measured $d\sigma /dt$
so that any approximation can easily be estimated.
The approximation sign refers to the neglected factor $1+\rho _0^2$. 
According to experimental data  $\rho _0 $(7 TeV, 0)$\approx 0.145$.
The parameter $\zeta $ is uniquely determined by the normalization of
the amplitude $f$. 

Thus, according to the unitarity condition the absorption for central 
collisions is governed by a single experimentally measured 
parameter $\zeta $ related to the share of elastic processes.
For central ($b=0$) collisions, the inelastic profile $G(s,0)$ achieves 
the maximum value equal to 1 (the full absorption) at $\zeta (s)=1$. 
It decreases parabolically $G(s,0)=1-\epsilon ^2$ for any decline of 
$\zeta $ ($\zeta =1\pm \epsilon $) from 1, i.e., it is very small 
for small decline $\epsilon $. The positivity of $G(s,0)$
imposes the limit $\zeta \leq 2$. At $\zeta =2$ the complete
transparency of central collisions $G(s,0)=0$ is achieved.

The elastic profile also reaches 1 at $b=0$ for $\zeta =1$ and
completely saturates the total profile for $\zeta =2$.

The experimentally measured share of elastic processes demonstrates 
non-trivial dependence on energy (see the Table in \cite{ufnel}). 
At low energies up to ISR the parameter $\zeta $ decreases (from about 1 down 
to values about 2/3) but then starts increasing and reaches its critical value 
1 for 7 TeV data at LHC. It is intriguing whether this increase will really 
show up in experiments at higher energies 
or it will be saturated asymptotically with $\zeta $ tending to 1 from below. 
The saturation would lead to the conservative stable situation of slow approach
to full absorption in central collisions while further increase will require 
the transition to another branch of the unitarity equation and new physics 
interpretation.

To explain these statements let us rewrite Eq. (\ref{gZ}) as
\begin{equation} 
 \zeta (s)=1\pm \sqrt {1-G(s,0)}.
\label{zeta}
\end{equation}
One used to treat elastic scattering as a shadow of inelastic processes. This 
statement is valid when the branch with negative sign is considered because it 
leads to proportionality of elastic and inelastic contributions for small 
$G(s,0)\ll 1$. That is typical for electrodynamics (e,g., for processes like
$ee\rightarrow ee\gamma$) and for optics. Therefore the increase of the elastic 
share at diminishing role of inelastic production came as a surprise. However,
for strong interactions, this share is close to 1 (see the Table). The approach 
of $\zeta $ to 1 at 7 TeV corresponds to complete absorption in central 
collisions. This value is considered as a critical one because from (\ref{zeta}) 
one gets significant conclusion that the excess of $\zeta $ over 1 implies that
the unitary branch with positive sign is at work. This branch was first 
considered in \cite{trt} with application to high energy particle scattering.
That changes the interpretation of the role of elastic processes as being 
a simple "shadow" of inelastic ones.

Some slight trend of $\zeta $ to increase and become larger than 1 can be 
noticed from comparison of TOTEM data at 7 TeV \cite{totem1} where it can 
be estimated\footnote{The experimental values of the ratios of elastic to total 
cross section and $\rho _0$ have been used.} in the limits 1.00 and 1.02 and 
at 8 TeV \cite{totem2} where it is about 1.04 though within the accuracy of 
experiments about $\pm $0.024. The precise data at 13 TeV are needed.
The further increase of the share of elastic scattering with energy is favored 
by extensive fits of available experimental information for the wide energy 
range and their extrapolations to ever higher energies done in Refs 
\cite{kfk, fms} as well as by some theoretical speculations (e.g., see Ref. 
\cite{roy}). The asymptotical values of $\zeta $ are about 1.5 in Refs
\cite{kfk, fms} and 1.8 \cite{roy}. They correspond to incomplete but
noticeable transparency at the center of the interaction region.

\section{The shape of the inelastic interaction region}

The detailed shape of the inelastic interaction region can be obtained with 
the help of relations (\ref{gamm}), (\ref{unit1}) if the behavior of the 
amplitude $f(s,t)$ is known. Its modulus and the $\rho _0$ values are
obtained from experiment. The most prominent feature of $d\sigma /dt $
is its rapid exponential decrease with increasing transferred momentum 
$\vert t\vert $, especially in the near forward diffraction cone. 
Inserting the exponential shape in Eqs (\ref{ze}), (\ref{gamm}) one can write
\begin{equation}
i\Gamma (s,b)\approx \frac {\sigma _t}{8\pi }\int _0^{\infty}d\vert t\vert 
\exp (-B\vert t\vert /2 )(i+\rho )J_0(b\sqrt {\vert t\vert }).
\label{gam2}
\end{equation}
Let us stress that the diffraction cone dominates the contribution to 
${\rm Re}\Gamma $ in Eqs ({gam2}), ({ze}) so strongly that the tail of the 
differential cross section at larger $\vert t\vert $ can be completely neglected 
at the level less than 0.1$\%$ by itself and it is suppressed additionally by 
the Bessel function $J_0$. The accuracy of the approximation was estimated 
using fits of the experimental differential 
cross section outside the diffraction cone by simplest analytical expressions.
Moreover, it was shown \cite{dnec, ads} by computing how well the versions 
with direct fits of experimental data and with their exponential approximation 
coincide if used in the unitarity condition.  Therefore the expression 
(\ref{gam2}) can be treated as following directly from experiment and
being very precise. Herefrom, one calculates
\begin{equation}
{\rm Re}\Gamma (s,b)= {\zeta }{\exp (-\frac {b^2}{2B})}.
\label{rega}
\end{equation}
Correspondingly, the shape of the inelastic profile for small $\rho _0$ is 
given by
\begin{equation}
G(s,b)=  \zeta \exp (-\frac {b^2}{2B})[2-\zeta \exp (-\frac {b^2}{2B})].
\label{ge}
\end{equation}
It depends on two measured quantities - the diffraction cone width
$B(s)$ and its ratio to the total cross section $\zeta $, and scales as a 
function of $b/\sqrt {2B}$. It has the maximum at
\begin{equation}
b_m^2=2B\ln \zeta
\label{bm}
\end{equation}
with maximum absorption $G(b_m)=1$ for $\zeta \geq 1$. For $\zeta <1$ (which is 
the case, e.g., at ISR energies) one gets incomplete absorption $G(s,b)<1$ at 
any physical $b\geq 0$ with the largest value reached at $b=0$ because the real 
maximum of $G$ would appear at non-physical values of $b$ for lower energies. 
Then the disk is semi-transparent.

At $\zeta =1$, which is reached at 7 TeV, the maximum is positioned exactly 
at $b=0$, and maximum absorption occurs there, i.e. $G(s,0)=1$. The disk center 
becomes black. The strongly absorptive core of the inelastic interaction region 
grows in size compared to ISR energies (see \cite{dnec}) as we see from 
expansion of Eq. (\ref{ge}) at small impact parameters:
\begin{equation}
G(s,b)= \zeta [2-\zeta -\frac {b^2}{B}(1-\zeta )-\frac {b^4}{4B^2}(2\zeta -1)].
\label{gb}
\end{equation}
The negative term proportional to $b^2$ vanishes at $\zeta =1$, and $G(b)$ 
develops a plateau which extends to quite large values of $b$ (about 0.5 fm). 
The plateau is very flat because the last term starts to play a role at 7 TeV 
(where $B\approx 20$ GeV$^{-2}$) only for larger values of $b$. 

With further increase of elastic scattering, i.e., at $\zeta >1$, the maximum 
shifts to positive physical impact parameters. A dip is formed at $b$=0 leading 
to a concave shaped inelastic interaction region - approaching a toroid-like 
shape (see \cite{jetp14, drem2, anis}). This dip becomes deeper at larger 
$\zeta $. The limiting value $\zeta =2$ leads to complete transparency at the 
center $b=0$ as discussed in the previous section. It can be only reached if the 
positive sign branch of the unitarity condition is applicable.

All these features are demonstrated in Fig. 1 borrowed from Ref. \cite{igse}.
\begin{figure}
\caption{ The evolution of the inelastic interaction region in terms 
of the survival probability. The values $\zeta =0.7$ and $1.0$ correspond to 
ISR and LHC energies and agree well with the result of detailed fitting to the 
elastic scattering data \cite{amal, dnec, mart}. A further increase of 
$\zeta $ leads to the toroid-like shape with a dip at $b=0$.
The values $\zeta =1.5$ are proposed in \cite{kfk, fms} and $\zeta =1.8$ in
\cite{roy} as corresponding to asymptotical regimes. The value $\zeta =2$
corresponds to the "black disk" regime ($\sigma _{el}=\sigma _{in}=
0.5\sigma _{tot}$). For more discussion of the black disk and the geometrical
scaling see Refs \cite{dias, csor, fag}.}
\centerline{\includegraphics[width=\textwidth, height=9cm]{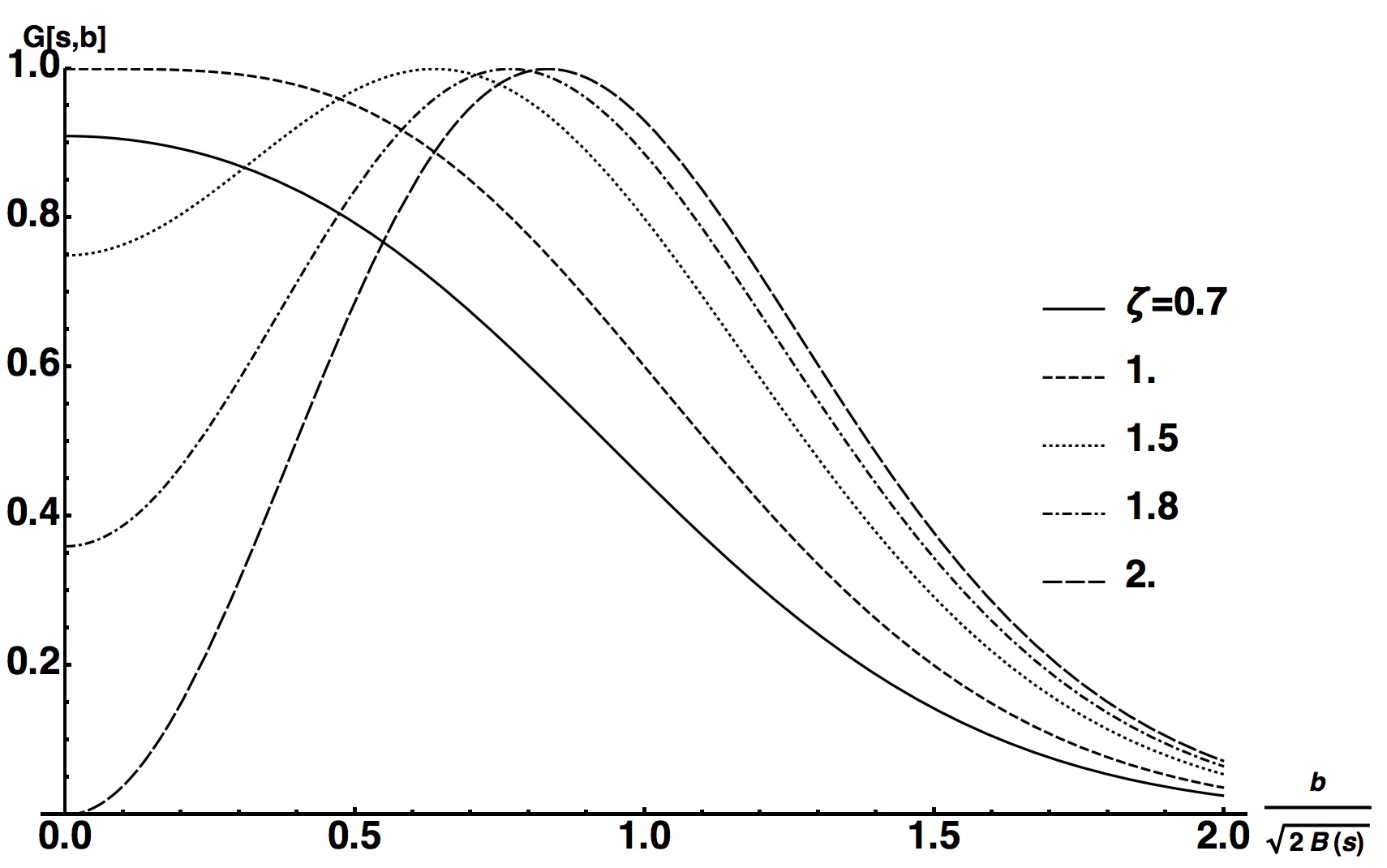}}
\end{figure}
The asymptotical regimes with further increase of the share of elastic 
scattering proposed in Refs \cite{kfk, fms, roy} predict the diminished 
absorption for central collisions. The whole space structure reminds the toroid 
(tube) with absorbing black edges which looks as if being more and more
transparent for the elastic component at the very center.  However,
the realistic estimates of its effects at the energies 13 TeV and 100 TeV
\cite{ksf} show that extremely high accuracy of experiments will be necessary
to observe these effects.

That is especially true because the cross sections of processes with small 
impact parameters are very small. Integrating the total and elastic terms in 
Eq. (\ref{ge}) up to impact parameters $b\leq r$ one estimates their roles 
for different radii $r$.
\begin{equation}
\sigma _{el}(s, b\leq r)=\sigma _{el}(s)[1-\exp (-r^2/B(s))],
\label{el}
\end{equation}
\begin{equation}
\sigma _{tot}(s, b\leq r)=\sigma _{tot}(s)[1-\exp (-r^2/2B(s))].
\label{tot}
\end{equation}
One gets that the contribution of processes at small impact parameters 
$b^2\ll 2B$ diminishes quadratically at small $r\rightarrow 0$. In particular,
inelastic processes contribute at $r\rightarrow 0$ as
\begin{equation}
\sigma _{in}(s, b\leq r)\rightarrow \pi r^2G(s,0)+O(r^4);\;\; (r^2\ll B).
\label{in}
\end{equation}
The maximum intensity of central collisions is at $\zeta =1$. That has been 
used in Ref. \cite{ads} for explanation of jets excess in very high multiplicity 
events at 7 TeV as an indication on the active role of gluons at that energy. 
It tends to 0 for $\zeta \rightarrow 2$. Thus, one predicts the diminished
role of jet production from central collisions with increase of $\zeta $.
It would ask for extremely precise data to reveal any evolution of that effect
at higher energies because according to estimates of Refs \cite{kfk, fms}
the decline from criticality is very small up to 100 TeV: 
$\zeta (13 TeV)=1.05-1.06;\;\;\; \zeta (95 TeV)=1.12-1.15$.

The peripheral regions dominate, especially in inelastic processes.

The spatial region of elastic scattering as derived from the subtrahend in
Eq. (\ref{ge}) is strongly peaked in the forward direction. The contribution to 
the elastic cross section is suppressed at small $b$ and comes mainly from 
impact parameters $b^2\approx 2B$. The average value of the squared impact
parameter for elastic scattering can be estimated as
\begin{equation}
<b_{el}^2>=\sigma _{el}(s)/\pi \zeta ^2(s).
\label{b2el}
\end{equation}
Inelastic processes are much more peripheral. The ratio of the corresponding
values of squared impact parameters is
\begin{equation}
\frac {<b_{in}^2>}{<b_{el}^2>}=\zeta \frac {8-\zeta }{4-\zeta }.
\label{in/el}
\end{equation}
This ratio exceeds 2 already at LHC energies and would become equal to 6
for (would be!) $\zeta =2$.  The peripherality of inelastic processes
compared to elastic ones increases with increase of the share of elastic 
collisions.

\section{Discussion and conclusions}

The intriguing increase of the share of elastic processes to the total outcome
observed at energies from ISR to LHC attracts much attention nowadays.
Its approach to 1/4 at LHC can become a critical sign of the changing character
of processes of proton interactions if the above tendency of increase persists.
The concave central part of the inelastic interaction region would be formed. 
The inelastic interaction region looks like a toroid hollowed inside and
strongly absorbing in its main body at the edges. The role of elastic
scattering in central collision becomes increasing. That is surprising and
contradicts somewhat to our theoretical prejudices. From the 
formal theoretical point of view it requires to consider another branch of
the unitarity condition that asks for its physics interpretation.

It is hard to believe that protons become more penetrable at higher energies
after being so dark in central collisions with $G(s,0)=1$ at 7 TeV unless
some special coherence within the internal region develops. Moreover, it
seems somewhat mystifying why the coherence is more significant just for 
central collisions but not at other impact parameters where inelastic 
collisions become dominant. The role of string junction in fermionic hadrons 
can become crucial. The relative strengths of the longitudinal and transverse
components of gluon (string) fields can probably explain the new physics
of hollowed hadrons.

One could imagine another classical effect that "black" protons start scattering 
in the opposite direction \cite{igse} like the billiard balls for head-on
collisions. Snell's law admits such situation for equal reflective indices of
colliding bodies. That can be checked if forward 
and backward scattered protons can be distinguished in experiment. Then they 
should wear different labels. One could use the proton spin as such a label. 
In principle, experiments with polarised protons can resolve the problem.
However, the more realistic classical scenario in this case would be complete
breaking the balls into pieces, i.e. dominance of inelastic processes.

Another hypothesis \cite{bj} treats the hollowed internal region as resulting 
from formation of cooler disoriented chiral condensate inside it ("baked-alaska" 
DCC). The signature of this squeezed coherent state would be some 
disbalance between the production of charged and neutral pions \cite{andr} 
noticed in some cosmic ray experiments. However the cross sections for central 
collisions seem to be extremely small as discussed above. The failure to find 
such events at Fermilab is probably connected with too low energies available.
It leaves some hope for higher energies in view of discussions above. Total
internal reflection of coherent states from dark edges of the toroid can be 
blamed for enlarged elastic scattering (like transmission of laser beams 
in optical fibers).

The transition to the deconfined state of quarks and gluons in the central 
collisions could also be claimed responsible for new effects (see Ref. 
\cite{trty}). The optical analogy with the scattering of light on metallic 
surface as induced by the presence of free electrons is used. Again, it is hard 
to explain why that happens for central collisions while peripheral ones with 
impact parameters near $b_m$ are strongly inelastic.

The last, but not the least, is the hypothesis that centrally colliding protons 
at $\zeta =2$ remind solitons which "pass through one another without losing
their identity. Here we have a nonlinear physical process in which
interacting localized pulses do not scatter irreversibly"\cite{krus}.
Non-linearity and dispersive properties (the chromopermittivity \cite{cher})
of a medium compete to produce such effect.

To conclude, the problem of increasing elastic cross section can be only solved
by experiment at higher energies. If this tendency persists, one should
invent new ways of explaining the transition to quite uncommon regime of
proton interactions with peculiar shapes of the interaction region.

\medskip 
{\bf Acknowlegment} 
 
I am grateful for support by the RFBR-grant 14-02-00099 and 
the RAS-CERN program.


\begin{thebibliography}{99}
\bibitem{totem1}
TOTEM Collaboration, {\it Nucl. Phys. B} {\bf 899}, 527 (2015).
\bibitem{totem2}
TOTEM Collaboration, {\it Phys. Rev. Lett.} {\bf 111}, 012001 (2013).
\bibitem{ufnel}
I.M. Dremin, {\it Physics-Uspekhi} {\bf 56}, 3 (2013).
\bibitem{drem2}
I.M. Dremin, {\it Physics-Uspekhi} {\bf 58} 61 (2015). 
\bibitem{jetp14}
I.M. Dremin, {\it JETP Lett.} {\bf 99}, 283 (2014).
\bibitem{igse}
I.M. Dremin and S.N. White, The interaction region of high energy protons; 
arXiv 1604.03469.
\bibitem{PDG}
PDG group, {\it China Phys. C} {\bf 38}, 090513 (2014).
\bibitem{anddre1}
I.V. Andreev and I.M. Dremin, {\it ZhETF Pis'ma} {\bf 6}, 810 (1967).
\bibitem{trt}
S.M. Troshin and N.E. Tyurin, {\it Phys. Lett. B} {\bf 316}, 175 (1993).
\bibitem{kfk}
A.K. Kohara, E. Ferreira and T. Kodama, {\it Eur. Phys. J. C} {\bf 74}, 3175 
(2014).
\bibitem{fms}
D.A. Fagundes, M.J. Menon and P.V.R.G. Silva, {\it Nucl. Phys. A} {\bf 946},
194 (2016).
\bibitem{roy}
S.M. Roy, A two component picture for high energy scattering: unitarity, 
analitycity and LHC data; arXiv:hep-ph, 1602.03627
\bibitem{dnec}
I.M. Dremin and V.A. Nechitailo, {\it Nucl. Phys. A} {\bf 916}, 241 (2013).
\bibitem{ads}
M.Yu. Azarkin, I.M. Dremin and M. Strikman, 
{\bf Phys. Lett. B} {\bf 735}, 244 (2014).
\bibitem{anis}
V.V. Anisovich, {\it Physics-Uspekhi} {\bf 58}, 1043 (2015).
\bibitem{amal}
U. Amaldi and K.R. Schubert, {\it Nucl. Phys. B} {\bf 166}, 301 (1980).
\bibitem{mart}
A. Alkin, E. Martynov, O. Kovalenko and S.M. Troshin, {\it Phys. Rev. D} 
{\bf 89}, 091501 (2014) (R).
\bibitem{dias}
P Brogueira and J Dias de Deus, {\it J. Phys. G} {\bf 39}, 055006 (2012)
\bibitem{csor}
T Cs"org"o and F Nemes, {\it IJMP A} {\bf 29}, 1450019 (2014)
\bibitem{fag}
D.A. Fagundes, M.J. Menon and P.V.R.G. Silva, {\it J. Phys. G} {\bf 40}, 065005
(2013)
\bibitem{ksf}
I.M. Dremin, {\it Bull. Lebedev Phys. Inst.} {\bf 42}, 8 (2015).
\bibitem{bj}
G. Amelino-Camelia, J.D. Bjorken and S.E. Larsson, {\it Phys. Rev. D} {\bf 56} 
, 6942 (1997).
\bibitem{andr}
I.V. Andreev, {\it Pisma v ZhETP} {\bf 33}, 384 (1981).
\bibitem{trty}
S.M. Troshin and N.E. Tyurin, {\it Mod.Phys.Lett. A} {\bf 31}, 1650025 (2016).
\bibitem{krus}
N.J. Zabusky and M.D. Kruskal, {\it Phys. Rev. Lett.} {\bf 15}, 240 (1965).
\bibitem{cher}
I.M. Dremin, {\it Nucl. Phys. A} {\bf 767}, 233 (2006).
\end{thebibliography}
\end{document}